# Topological Nodal Chains and Transverse Transports in Ferromagnetic Centrosymmetric Semimetal FeIn$_2$S$_4$

Junyan Liu[1], Yibo Wang[1,2], Xuebin Dong[1,2], Jinying Yang[1,2], Shen Zhang[1,3], Meng Lyu[1], Binbin Wang[1], Hongxiang Wei[1], Shouguo Wang[4,5,*], Enke Liu[1,*], Baogen Shen[1,3]

[1]Beijing National Laboratory for Condensed Matter Physics, Institute of Physics, Chinese Academy of Sciences, Beijing 100190, China
[2]School of Physical Sciences, University of Chinese Academy of Sciences, Beijing 100049, China
[3]Ningbo Institute of Materials Technology & Engineering, Chinese Academy of Sciences, Ningbo, Zhejiang 315201, China
[4]School of Materials Science and Engineering, Anhui University, Hefei 230601, China
[5]School of Materials Science and Engineering, University of Science and Technology of Beijing, Beijing, 100083 P. R. China

*E-mail: sgwang@ahu.edu.cn, ekliu@iphy.ac.cn

**Abstract**

Nodal chain semimetals protected by nonsymmorphic symmetries are distinct from Dirac and Weyl semimetals, featuring unconventional topological surface states and resulting in anomalous magnetotransport properties. Here, we reveal that the ferromagnetic FeIn$_2$S$_4$ is a suitable nodal chain candidate in theory. Centrosymmetric FeIn$_2$S$_4$ with nonsymmorphic symmetries shows half-metallicity and clean band-crossings with hourglass-type dispersion tracing out nodal lines. Owing to glide mirror symmetries, the nontrivial nodal loops form nodal chain, which is associated with the perpendicular glide mirror planes. These nodal chains are robust against spin-orbit interaction, giving rise to the coexistence of drumhead-type surface states and closed surface Fermi arcs. Moreover, the nodal loops protected by nonsymmorphic symmetry contribute to large anomalous Hall conductivity and the anomalous Nernst conductivity. Our results provide a platform to explore the intriguing topological state and transverse transport properties in magnetic system.

# I. INTRODUCTION

Recently discovered topological nodal-line metals [1–4] and topological semimetals including Dirac and Weyl semimetals [5–9], have been concerned by more and more researchers because it hosts robust low-energy fermionic excitations. The low-energy excitations described by Dirac or Weyl Hamiltonians are different from the electrons in conventional materials, such as high mobility [10], opposite chirality [11] and topological robustness [12], and give rise to many fascinating properties in topological metals and semimetals [3,13]. For example, the Weyl fermions in non-magnetic materials present topological surface Fermi arcs [14,15], and the magnetic Weyl semimetals lead to the realization of the chiral anomaly and Fermi arcs [16,17]. The special 'drumhead' surface states [7,18–20] and the anomalous Landau level spectrum [21] have been proposed in nodal-line metals whose nodes extend along one-dimensional (1D) lines instead of discrete points in the three-dimensional (3D) Brillouin zone (BZ). However, such 1D nodal loops (closed nodal line) protected by certain symmorphic symmetry operations, such as mirror or inversion [22–25], are usually vulnerable against spin-orbit coupling (SOC), and can be removed without altering the symmetry [26]. There is a series of works on the absence of extended degeneracies with SOC in different space groups. For example, the fourfold double Weyl points of orthorhombic crystal $Ag_2Se$ in SG 19 become twofold degenerate under SOC [27].

Compared with the symmorphic symmetry operations, the non-symmorphic glide-plane symmetry $g = \{\sigma/t\}$, formed by a reflection $\sigma$, followed by a translation by a fraction of a primitive lattice vector, $t$, plays a critical role in stabilizing the band-crossing points [1,4,28–31]. This band-crossing points are robust against SOC and may entangle multiple bands together, resulting in inevitable crossing points. The entangled multiple nodal loops can form nodal chain configurations including the outer nodal chain (two nodal lines are on opposite sides of the touching point) [2,4,28], inner nodal chain (two nodal lines are on same sides of the touching point) [32], as shown in Fig.1, and Hourglass Dirac chain (the four-fold degenerate neck crossing-point) [4,33–35]. A variety of outer nodal chain metals proposed in non-centrosymmetric system always enclose a time-reversal invariant momenta (TRIM), which was predicted in paramagnetic $IrF_4$ material [1]. Furthermore, $IrF_4$ also exhibits an antiferromagnetic semimetal with the nodal line [36]. The hourglass Dirac chain metal dictates by two orthogonal glide mirror planes combined with 𝒯 and 𝒫 symmetries, which was predicted in the rhenium dioxide [4]. Moreover, the outer and inner nodal chains can coexist in ferromagnetic Heusler $Co_2MnGa$ material [2]. Recently, the outer nodal chain in a metallic-mesh photonic crystal and Dirac nodal chain in a layered structure centrosymmetric $TiB_2$ have been observed by the ARPES measurements [19]. So far, the proposed inner nodal chain semimetals are still limited, and it is urgent to discover more suitable candidates to explore their intriguing properties.

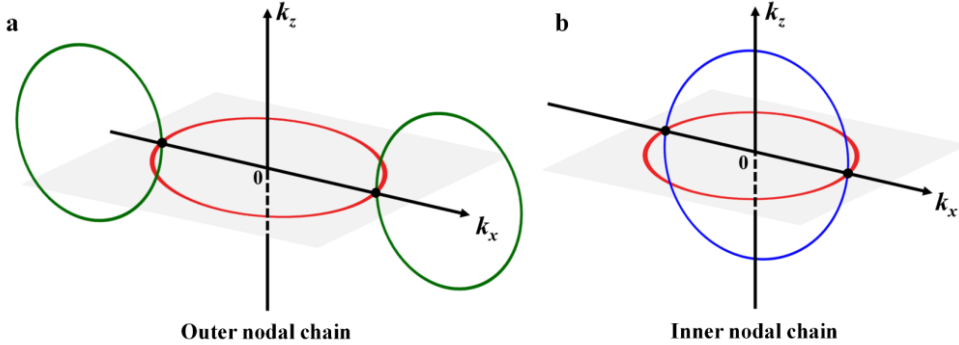

**Fig. 1**. Schematic figure of the (a) inner and (b) outer nodal chains.

In this work, using first-principles calculations, we predict the ferromagnetic centrosymmetric semimetal $FeIn_2S_4$ as a candidate of inner nodal chain material. Ferromagnetic $FeIn_2S_4$ intrinsically shows the half-metallic feature and clean band-crossing with high fermi velocity near Fermi level. The effective model and symmetry analysis combining with calculations reveal that $FeIn_2S_4$ is characterized by inner nodal chains, which are protected by orthogonal glide mirror planes. The band crossing points trace out nodal loops and the connected nodal loops form inner nodal chains in the momentum space, being robust against SOC. The coexistence of drumhead-type surface states and closed surface Fermi arcs are observed by the surface Green's function method. Moreover, these nontrivial nodal loops contribute to the intrinsic anomalous Hall conductivity and the anomalous Nernst conductivity. The result suggests that ferromagnetic $FeIn_2S_4$ is a potential candidate for exploring the intriguing topological state and transverse transport properties in a magnetic system.

## II. METHOD

First-principles calculations based on the density-functional theory (DFT) were performed using the projector-augmented wave (PAW) method [37] as implemented in the Vienna Ab

Initio Simulation (VASP) Package [38]. The generalized gradient approximation (GGA) of Perdew-Burke-Ernzerhof (PBE) [39] for the exchange-correlation functional was used. The cut-off energy of plane wave basis was set 600 eV and the first Brillouin zone of the reciprocal space was sampled with Monkhorst-Pack $k$-point meshes [40] of 11×11×11 for structural relaxation and 15×15×15 for static self-consistent calculation. The structures were fully relaxed until the force and energy is smaller than 0.001 eV/Å and $10^{-6}$ eV, the optimized lattice constant is 10.694 Å for FeIn$_2$S$_4$. The band representation analysis was performed using the Spacegroup package [41]. The Fe $d$, In $s$, and S $p$ orbitals were used to construct the tight-binding model with the maximally localized Wannier functions (MLWF) by Wannier90 code [42]. Energy dependence of the anomalous Hall conductivity (AHC) $\sigma_{xy}^z$ in terms of the $z$ components of Berry curvature was obtained by the WannierBerri code [43]. The anomalous transverse thermoelectric conductivity $\alpha_{xy}^A$ was obtained from the integral formula of Mott relation [44]. The calculation of topological surface spectrum and surface states are based on the surface Green's function method as implemented in the Wanniertools package [45].

**III. CRYSTAL STRUCTURE, SYMMETRY, AND MAGNETIC CONFIGURATION**

The crystal structure of FeIn$_2$S$_4$, synthesized by the method of directional crystallization of an almost stoichiometric melt and obtained bulk single crystal in experiment [46], belongs to face-centered cubic lattice and is composed of Fe-S tetrahedrons and In-S octahedrons, with each Fe atom surrounded by four sulfur atoms and each In atom surrounded by six sulfur atoms, as shown in Fig. 2a. Without spin-orbit coupling (SOC), spin rotations and the symmetry operations of lattice can be a combination of a spatial operation and an arbitrary spin rotation that is compatible with the group structure, classifying spin group as spin space group (SSG) [47] and spin point group (SPG) [48]. Here, the symmetries of space and spin degrees of freedom are considered separately. The space group F$d$-3$m$ (No. 227) symmetry of the FeIn$_2$S$_4$ structure includes the following nonsymmorphic symmetry operations: two glide mirror planes with a fraction translations $\widetilde{\mathcal{M}}_z: (x, y, z) \to \left(x + \frac{1}{4}, y + \frac{3}{4}, -z + \frac{1}{2}\right)$, $\widetilde{\mathcal{M}}_{xy}: (x, y, z) \to \left(-y + \frac{1}{4}, -x + \frac{3}{4}, z + \frac{1}{2}\right)$. The corresponding first Brillouin zone (BZ) and 2D BZ for (001) surface are shown in Fig. 2b. To obtain the magnetic ground state of FeIn$_2$S$_4$, the total energies for nonmagnetic (NM), ferromagnetic (FM) (only give (001) direction due to isotropy) and antiferromagnetic (AFM, possible AFM configurations in Fig.S1, see the Supplemental Material [49]) are calculated and the result reveals that the FM state is the most stable energetically, as shown in Table I. Notably, $FeIn_2S_4$ shows an antiferromagnetic state in experiment [50], in which the magnetic moments are arranged in spin-antiparallel collinearly [51]. The difference between experiment and calculation may due to small strain during experimental preparation, such as stress, defect [52] or disorder, which make $FeIn_2S_4$ fail to adopt the FM state. From our calculations, the total energy of FM state is very close to that of AFM state and the spin-flip transition from FM state to AFM state can be achieved by only 1% tensile strain in Fig. 2c. If ferromagnetic $FeIn_2S_4$ can been obtained by experiment, it will show interesting topological state, being discussed in the section V. Therefore, it deserves a deep theoretical exploration on the potential FM state and the striking nontrivial topological states in $FeIn_2S_4$ compound. For FM state, each Fe atom has a nominal magnetic moment of 4 $\mu_B$. According to the Pauling electronegativity, the value of electronegativity of S atom (2.58) is greater than that of the In atom (1.78) and Fe atom (1.83), resulting in two In atoms donating six electrons from $5p$ and $5s$ orbitals and Fe atom donating two electrons from $4s$ orbital to S-$3p$ orbital in Fig. 2d. As a consequence, the S-$3p$ orbital is occupied by two spin-antiparallel electrons, exhibiting nearly zero magnetic moment. According to the eight-electron rule, the Fe atoms would completely hold six valence electrons, leaving four unpaired electrons with a magnetic moment of 4 $\mu_B$.

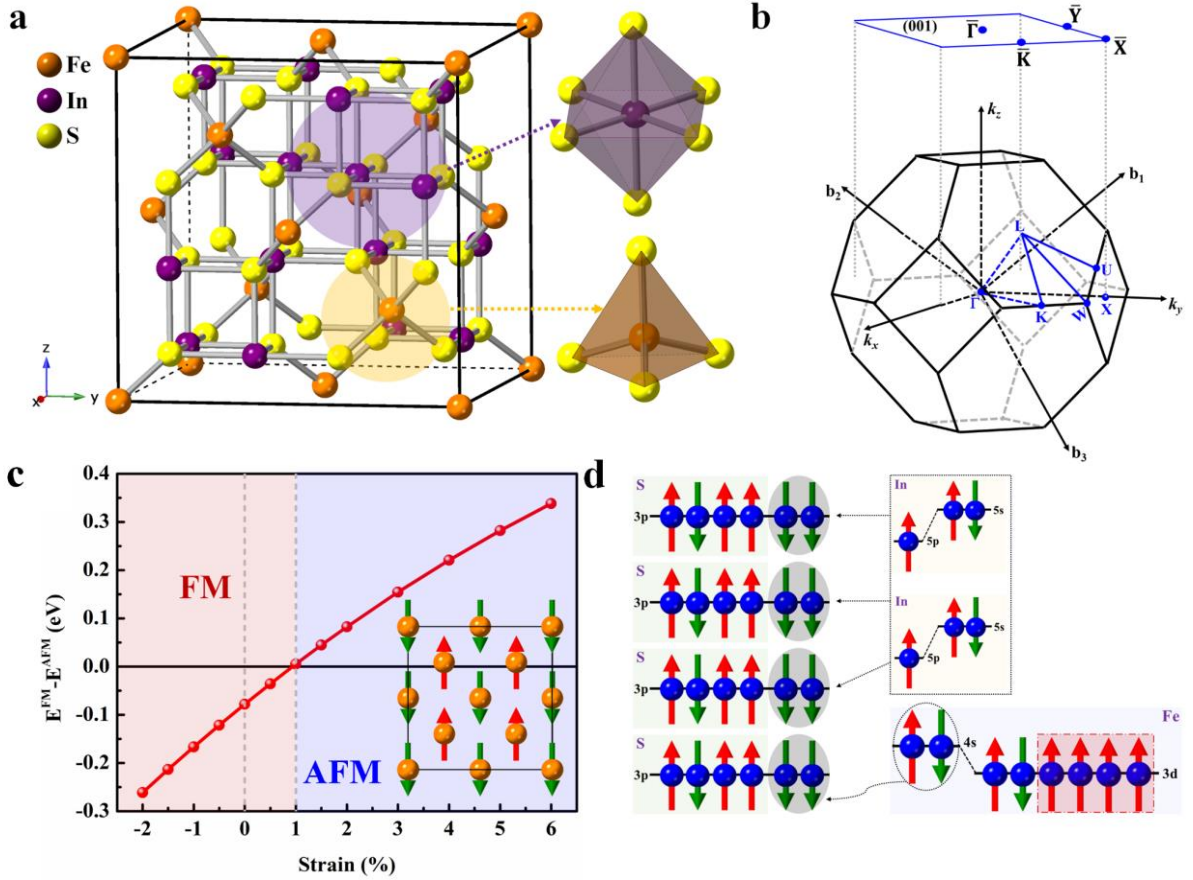

**Fig. 2.** (a) Structure of FeIn$_2$S$_4$ with space group *Fd-3m* (No. 227). The In-S octahedron and Fe-S tetrahedron structure units are displayed in the top right and bottom right, respectively. (b) Three-dimensional (3D) Brillouin zone (BZ) and 2D BZ for (001) surface. (c) Strain dependence of the difference of the total energy between FM and AFM configurations. (d) Schematic diagram for the origin of the nominal magnetic moment of Fe atoms, 4 $\mu_B$ per Fe atom in the FM configuration.

**TABLE I**. Total energy per unit cell of FeIn$_2$S$_4$ with NM, AFM and FM configuration (unit in eV).

|  | FM | AFM | NM |
| --- | --- | --- | --- |
| Total energy | -66.084 | -66.035 | -63.967 |

## IV. ELECTRONIC STRUCTURE

The spin-resolved band structure of FM FeIn$_2$S$_4$ exhibits half-metallic nature because the spin-up channel shows a semiconductor behavior, while the spin-down channel displays a metallic property, where the conducting electrons have a 100% spin polarization, as shown in Fig. 3a-b. The spin-down channel mainly attributes to Fe-*d* orbitals, being similar to that of Co$_3$Sn$_2$S$_2$ [9]. Significantly, for spin-down channel, the highest valence bands cross the lowest conduction bands linearly along Γ-X, Γ-K, L-U and L-K high symmetry lines, of which the band crossing along Γ-X direction is far away from Fermi level. The band along W-L direction opens a tiny band gap though it shows a linear band crossing visually. Figure 3c presents the enlarged plot of linear dispersion between energy and momentum near the Fermi level for the spin-down channel. The band-crossing points locating at X-Γ-K-L-U lines are referred to as BC1, BC2, BC3 and BC4, respectively. BC3 and BC4 are symmetrically distributed in K-L-U lines. The linear dispersion indicates high carrier mobility and the excellent performance of a half metal. From the two linear bands highlighted in Fig. 3c, the Fermi velocity $v_\mathrm{F}$ of the carriers can be evaluated using linear fitting [53]: $v_\mathrm{F} \approx \frac{1}{\hbar}\frac{\partial E}{\partial k}$. The Fermi velocity in Fig. 3d reveal that: i) half metal $FeIn_2S_4$ possesses high carrier mobility with Fermi velocity of about $2\times10^5$ m/s; ii) double degenerate points along Γ-K, K-L and L-U directions are inclined Weyl cones. Moreover, according to the band representation analysis of the symmetric operations, the band crossing points locating at Γ-X and Γ-K lines are P-WNLs, i.e. the Weyl nodal-line net (WNL net) contains multiple two-fold NLs, which share (at least) one nodal point in momentum space, and then the joint nodal point are termed as P-WNLs. The Quadratic contact triple points (QCTP, i.e. three-fold band degeneracy) at Γ point are ignored due to the topological charge $C = 0$ [54].

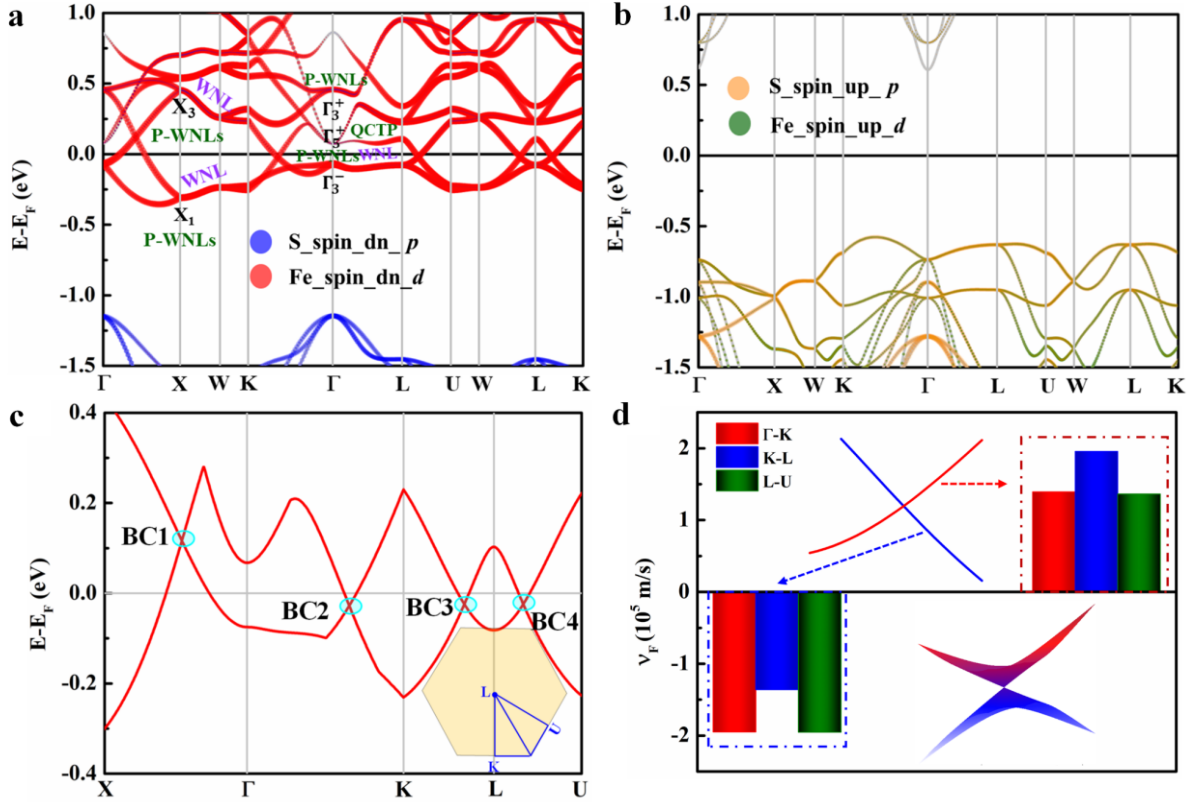

**Fig. 3.** (a)-(b) Orbital-projected spin-polarized bands structure of FeIn$_2$S$_4$, Fe-$d$ orbitals mainly contribute to the band near Fermi level for spin-down channel, band representation and topological state at Γ and X points from symmetry analysis. (c) The enlarged band structure around the two-fold-degenerate band-crossing points near Fermi level for spin-down channel. (d) Fermi velocity of the band crossings along Γ-K, K-L and L-U directions.

## V. NODAL CHAINS WITHOUT AND WITH SOC

To further analyze the topological character of FeIn$_2$S$_4$ suggested by Fig. 3c, the linear band crossings are traced out in more details. Without SOC, the spin-up and spin-down channels are decoupled, and thus the time-reversal symmetry can translate between the two spin channels. For the spin-down channel, all crystalline symmetries are preserved, such as nonsymmorphic glide mirror symmetries, inversion symmetry $\mathcal{P}$ and time-reversal symmetry $\mathcal{T}$. Figure 4a displays two kinds of nodal lines, in which the red nodal loops at $k_x = 0$ ($k_y = 0$ and $k_z = 0$) plane are protected by glide mirror plane $\widetilde{\mathcal{M}}_z(\widetilde{\mathcal{M}}_x, \widetilde{\mathcal{M}}_y)$, the black green nodal loop at $k_{x-z}$ plane is protected by mirror plane M$_z$. Remarkably, these nodal loops are interconnected and the red nodal loops are connected to red and black green ones to form inner nodal chains. At $k_{x-y}$ plane, the nodal lines throughout the Brillouin Zone from top to bottom are protected by $\widetilde{\mathcal{M}}_{xy}$ in Fig. 4b. These nodal loops can be argued from effective model and symmetry analysis. The presence of red nodal loops centered at the X point are protected by glide mirror plane $\widetilde{\mathcal{M}}_z: (k_x, k_y, k_z) \to \left(k_x + \frac{1}{4}, k_y + \frac{3}{4}, -k_z + \frac{1}{2}\right)$ and are caused by band inversion. To protect this nodal loop, the two crossing bands should have opposite eigenvalues of $\widetilde{\mathcal{M}}_z$, which has a matrix representation as $\widetilde{\mathcal{M}}_z = \sigma_z$. The two-band $k \cdot p$ Hamiltonian is given by

$$\mathcal{H} = \sum_{i=x,y,z} d_i(\mathbf{k}) \sigma_i, \tag{1}$$

where $\sigma_i$ is the Pauli matrix denoting the space of the two crossing bands, one with positive parity and the other with negative parity and $d_i(\mathbf{k})$ are real functions and the vector $\mathbf{k}$ is relative to the Γ point. To satisfy the commutation relation between $\widetilde{\mathcal{M}}_z$ and $\mathcal{H}$,

$$\widetilde{\mathcal{M}}_z \mathcal{H} \widetilde{\mathcal{M}}_z^{-1} = \mathcal{H}_\Gamma(k_x, k_y, -k_z), \tag{2}$$

which leads to $d_{x,y}(\mathbf{k})$ are even functions of $\mathbf{k}$ and $d_z(\mathbf{k})$ are odd functions of $\mathbf{k}$,

$$d_{x,y}(k_x, k_y, k_z) = -d_{x,y}(k_x, k_y, -k_z), \tag{3}$$

$$d_z(k_x, k_y, k_z) = d_z(k_x, k_y, -k_z). \tag{4}$$

When the nodal loops lie on $k_z = 0$ plane, $d_{x,y}(k_x, k_y, k_z) = -d_{x,y}(k_x, k_y, -k_z)$ vanishes, and thus the nodal loops are determined by $d_z(\mathbf{k})$. Considering the cubic symmetry and the second order of $\mathbf{k}$, $d_z(\mathbf{k})$ can be generally expressed in the following form:

$$d_z(\mathbf{k}) = m - b(k_x^2 + k_y^2 + k_z^2), \tag{5}$$

$d_z(\mathbf{k}) = 0$ can be satisfied only when $mb > 0$, indicating the nodal loops at $k_z = 0$ plane due to the band-inversion. Interestingly, the nodal loops at $k_x = 0$ and $k_y = 0$ plane are orthogonal to each other, and they touch and constitute inner chain in the momentum space (Fig. 4a). Interestingly, the nodal chains on perpendicular glide mirror planes are in contrast to the Hopf link, which represents 3D band crossings characterized by the simplest topologically nontrivial link and consists of two rings that pass through the center of each other [2], involving only two bands, and both glide mirror eigenvalues flip from one region to the other, as shown in Fig. 4c. The analysis of nodal lines (Fig. 4b) throughout the Brillouin Zone from top to bottom at $k_{x-y}$ plane protected by glide mirror plane $\widetilde{\mathcal{M}}_{xy}$ are similar.

At $k_{x-y}$ plane, the presence of black green nodal loops are protected by mirror plane

$\mathcal{M}_{x-y}: (x, y, z) \rightarrow (y, x, z)$. Since $\mathcal{M}_{1-10}: (k_x, k_y, k_z) \rightarrow (k_y, k_x, k_z)$ satisfies $(\mathcal{M}_{x-y})^2 = 1$ and then $\mathcal{M}_{x-y}$ has eigenvalues $\pm 1$.

It is worth noting that the presence of glide mirror symmetries results in a band switching between two time reversal invariant momenta (TRIM) Γ (0,0,0) and X (0, π, 0) points, forming the double-fold degeneracy hourglass type dispersion along Γ-X path (Fig. 4a). For the mirror-invariant line Γ-X on the $k_z = 0$ plane, since $\widetilde{\mathcal{M}}_z^4 = T_{310} = e^{-3ik_x - ik_y}$, and thus the eigenvalues of $\widetilde{\mathcal{M}}_z$ is $g_z = \pm 1$ at Γ point and $g_z = \pm i$ at X point. Without SOC, each spin channel can be regarded as a spinless case and the band dispersion are Kramer degeneracy at the Γ and X points. At Γ point, every Bloch state on plane $k_z = 0$ can be chosen as the eigenstate $|u\rangle$ of $\widetilde{\mathcal{M}}_z$; it shows a double degeneracy, $\{|u\rangle, \mathcal{T}|u\rangle\}$. For example, $|u\rangle$ have the same eigenvalue $g_z = +1$ (or $g_z = -1$) as its Kramer partner $\mathcal{T}|u\rangle$. Notably, at X = (0, π, 0) each Kramer pair $|u\rangle$ and $\mathcal{T}|u\rangle$ shares different eigenvalues $g_z$ (one is $g_z = i$ and the other one is $g_z = -i$). Therefore, there must be a partner switching when going from Γ to X, leading to an hourglass band crossing, as shown in Fig. 4d.

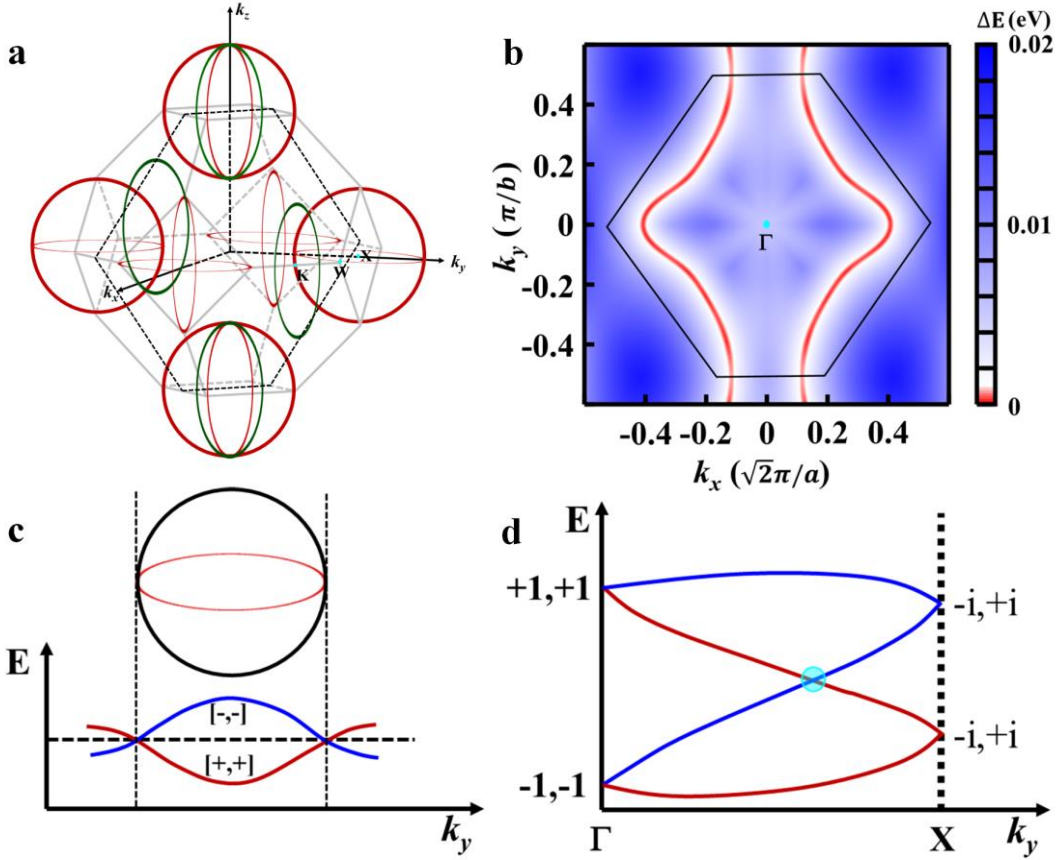

**Fig. 4.** (a) Schematic figure of two types of nodal loops forming inner nodal chains. (b) Energy gap ΔE

between the lowest conduction band and the top valence band without SOC at $k_{x-y} = 0$ plane, i.e. the nodal line throughout the Brillouin Zone from top to bottom. (c) Schematic figure of two independent nodal loops on two perpendicular glide mirror planes. The below panel shows the nodal loops with glide mirror eigenvalues along $k_y$ axis. (d) A nonsymmorphic symmetry $\{g|t\}$ leads to band crossings on a mirror-invariant line Γ-X. The Kramers pairs (Γ and X points) exchange the glide eigenvalues along any path connecting Γ and X points, leading to an essential crossing point.

When considering SOC effect that couples the two spins and allows them to hybridize, the above-mentioned band-crossings are not fully gapped, such as the crossing-points along Γ-X and Γ-K directions, as shown in Fig. 6a and Fig. S2[49]. Since all the glide mirrors and mirror symmetries except $\widetilde{\mathcal{M}}_z$, $\widetilde{\mathcal{M}}_{xy}$ and $M_{x-y}$ are broken under the magnetic moments of Fe atoms aligned in the (001) direction, the red nodal loops at $k_z = 0$, the green nodal loops at $k_{x-y}$ plane and the nodal lines throughout the Brillouin zone from top to bottom at $k_{-x-y}$ plane are retained. This illuminates that the band-crossing points along Γ-X and Γ-K directions enabled by nonsymmorphic symmetries are robust against SOC and referred to as essential. The red nodal loops at $k_x = 0$ and $k_y = 0$ are gapped and the SOC-induced gap opened at the nodal loops is extremely small of 1.3 meV. Therefore, the SOC effect is quite weak. Considering the negligible SOC gap, it is very likely that the inner nodal chains structure will be obtained in FeIn$_2$S$_4$. In addition, the crossing-points along L-U and L-K directions protected by mirror symmetry $M_{x-y}$ are fully gapped and the SOC-induced gaps are 19.8 meV and 27.7 meV, as shown in Fig. 6a.

**VI. DRUMHEAD SURFACE STATES**

For a topological nodal line material, there should exist surface states despite not having an exact bulk boundary correspondence, typically boundary modes are observed in IrF$_4$ [1], ReO$_2$ [4] and Mg$_2$VO$_4$[34]. Its topological surface spectrum along high symmetry paths in surface BZ have been established as the so-called drum-head surface states. A nodal line projected into a plane should fill the entire area inside the nodal line. The topological surface states fill the area shared by the two nodal lines for the nodal lines connecting with each other and forming nodal networks and the two drum-head surface states are coupled together. Interestingly, Nodal chain semimetals feature unconventional topological surface states. Figure 5a displays the surface spectrum near the Fermi level for the (001) plane (Fig. 2b) in FeIn$_2$S$_4$ with Fe termination. The sharp surface bands along $\overline{\Gamma}$-$\overline{X}$-$\overline{K}$ paths are observed and they actually are the drumhead surface states stemming from the projected bulk band-crossing point. Both the hourglass dispersion and the surface states are close to the Fermi level, indicating that they could be directly imaged by ARPES experiment in ferromagnetic FeIn$_2$S$_4$, and could dominate the electronic and thermal transport behaviors of the compound.

In Fig. 5b, the projected loops in the surface BZ for the constant energy slice at 50 meV below the Fermi level (Fig. 5a) are obtained by the tight-binding model. In order to clearly present the surface states of the nodal loops, the nodal loops outline the profiles projected onto the (001) surface in Fig. 5c. The opened circle surface states marked by the shaded region for the red nodal loop on the (001) surface are indeed observed, with the projected loops centering around X point. Other projected loops fill the cross-shaped area in the shaded region. An interesting coexistence of drumhead-type surface state and closed surface Fermi arcs are observed.

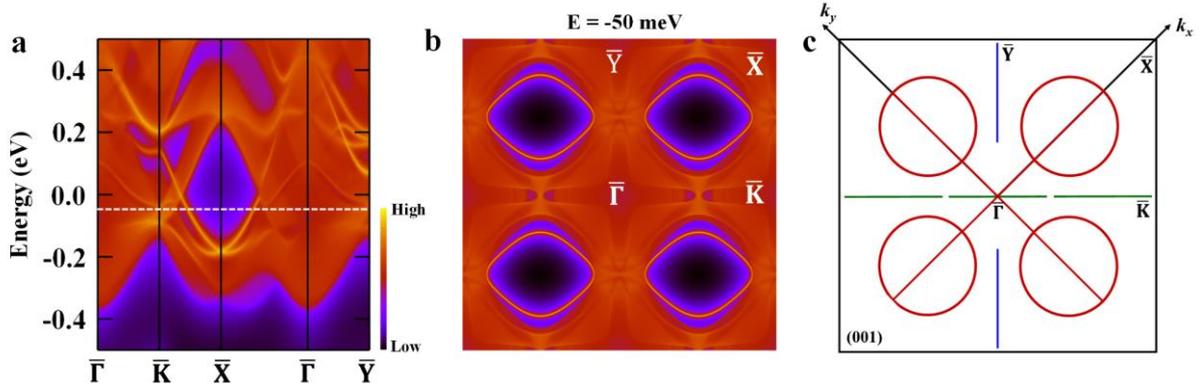

**Fig. 5.** (a) Topological surface states for (001) surface in FeIn$_2$S$_4$ with Fe termination. The sharp features are surface states and the shaded region are projections of bulk bands. (b) The projection of nodal features onto the (001) surface at 50 meV below the Fermi level. (c) Schematic illustration of projected topological nodal features observed in (b).

**VII. TRANSVERSE TRANSPORTS**

The inner chain is essential, robust against SOC, indicating that the interplay of magnetism and topology may open up the possibility for exotic linear response effects, such as the anomalous Hall effect and the anomalous Nernst effect. The energy-dependent anomalous Hall conductivity (AHC) and the anomalous Nernst conductivity (ANC) obtained from the Berry curvature [55] and the generalized Mott relation [44] $\alpha_{xy}^A(T,\mu) = -\frac{1}{e}\int d\epsilon \frac{\partial f(\epsilon-\mu,T)}{\partial \epsilon}\frac{\epsilon-\mu}{T}\sigma_{xy}^A(\epsilon)$ are shown in Figs. 6b-c. The gaped band crossing points along

U-L-K directions (Fig. 6a) correspond to a peak of 192 $(\Omega \cdot cm)^{-1}$ in $\sigma_{yx}$ at the energy of 15 meV below $E_F$ (Fig. 6b). The SOC-induced gapped nodal loops indeed create anomalous Hall effect. Considering energy shift away from the charge neutral point in a range of 250 meV, the maximum of $\sigma_{yx}$ -648 $(\Omega \cdot cm)^{-1}$ may appear at the energy of -83 meV owing to the possible vacancies, defects or off-stoichiometric composition in real materials. Energy-dependent anomalous Nernst conductivity in Fig. 6c shows that the maximum of $\alpha_{yx}$ can reach 2.0 $(A \cdot m^{-1} K^{-1})$ and -2.3 $(A \cdot m^{-1} K^{-1})$ at low temperature of 50 K, corresponding to the largest energy derivative of the AHC. Moreover, the maximum of $\alpha_{yx}$ are separated on either side of $\sigma_{yx}$, indicating that a small shift in the energy can also greatly enhance or decrease the ANC value and its sign change. The maximum of $\alpha_{yx}$ at energy of 15 meV below the Fermi level can trace back to the SOC-induced gapped nodal loops. As the temperature increases, the ANC changes from negative to positive value at the Fermi level (Fig. 6c), the low-temperature Mott relation is valid, and the $\alpha_{yx}$ can still maintain a high value of 2.30 $(A \cdot m^{-1} K^{-1})$ at 200 K. The maximum ANC of 2.0 $(A \cdot m^{-1} K^{-1})$ at T = 50 K is much larger than that in traditional ferromagnets [56] (typically $\alpha_{yx}$ = 0.01-1 $(A \cdot m^{-1} K^{-1})$) once T ≥ 50 K.

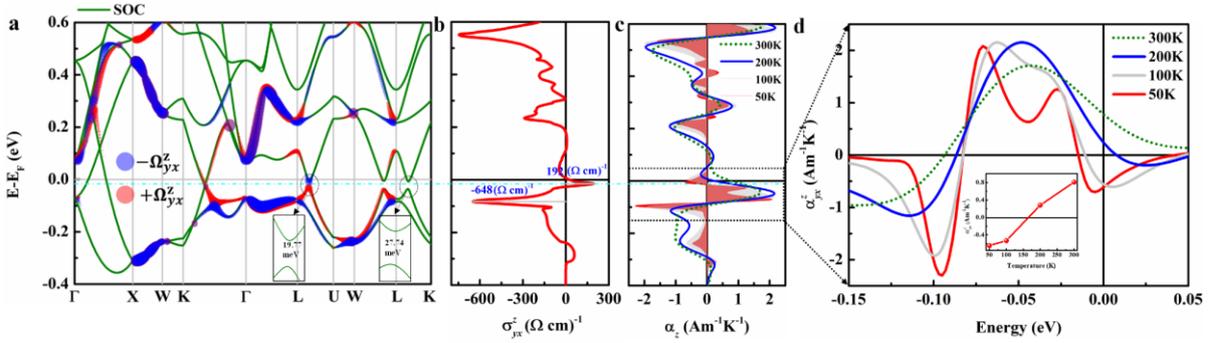

**Fig. 6.** (a) Energy dispersion of electronic bands structure of FeIn$_2$S$_4$ along high-symmetry paths in the Brillouin zone with SOC, berry curvature projected band structure along high-symmetry paths except L-K direction to highlight the SOC effect on the band-crossing, the size of the blue and red dots corresponds to the magnitude of negative and positive berry curvature on a logarithmic scale. (b) Energy dependence of the anomalous Hall conductivity in terms of the components of $\Omega_{yx}^z(\mathbf{k})$. (c) Energy and temperature dependence of the anomalous Nernst conductivity $\alpha_{yx}^z$ in terms of $\sigma_{yx}^z(\epsilon)$, right panel (d) are the enlarged view of ANC below the Fermi level, the illustration shows the curve of ANC at the Fermi level with temperature.

To investigate the origin of the anomalous Hall and Nernst effects, the integral Berry curvature $\Omega_{xy}^z(\mathbf{k})$ (BC) along $k_z$ in the Brillouin zone are obtained. First, the AHC of 192 $(\Omega \cdot cm)^{-1}$ can trace to the SOC-induced gapped nodal loop protected by $M_{x-y}$ in Fig. 6b.

Figure 7a shows that the distribution of the BC in the Brillouin zone that is focused only around the nodal loop, i.e. the integrated BC is primarily determined by the shape of the nodal lines, and the positive BC mainly contribute to the AHC. This is because the corresponding nodal lines along the nodal loop have less dispersive energy, while the dispersion nodal lines cause negative BC, which is the reason that maximum $\sigma_{yx}$ does not coincide with the crossing along the U-L-K line. Consequently, the gapped band crossing points exhibit a peak in AHC and ANC around the Fermi level. To find out the contribution of the maximum $\sigma_{yx}$, one can look into the BC distribution in the Brillouin zone. Figure 7b shows that the negative BC at $k_z = 0$ and $k_{x-y}$ planes, forming a nodal line in nodal loop case, are the dominating contribution to AHC of -648 $(\Omega \cdot cm)^{-1}$. Finally, we also examine other ferromagnetic and antiferromagnetic systems possessing nodal lines and nodal planes protected by nonsymmorphic symmetries and having similar transverse transport mechanisms. For example, the AHC of ferromagnetic $CuCr_2Se_4$ stems from a large Berry curvature due to the splitting of the nodal line via spin-orbit coupling [57]. The Berry-curvature "hot spots" lying along the gapped nodal lines via spin-orbit coupling enhance the anomalous Hall effect in collinear C-type antiferromagnetic $CaCrO_3$ [58]. The slightly gapped nodal plane enforced by a screw axis symmetry generates the spontaneous Hall and Nernst effects in compensated antiferromagnets $CoNb_3S_6$[59].

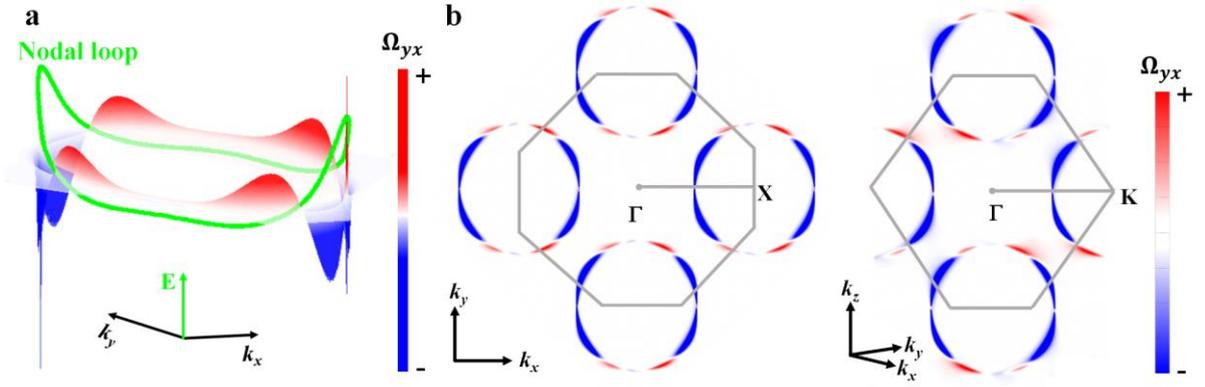

**Fig. 7.** (a) Berry curvature distribution in the Brillouin zone along the 3D band structure ($E$(eV) - $k_x$ ($2\sqrt{2}\pi/a$) - $k_y$ ($2\pi/b$)) of a green nodal loop at $k_{x-y}$ plane. (b) Berry curvature distribution projected to the $k_z = 0$ plane and $k_{x-y} = 0$ plane.

## VII. SUMMARY

In summary, the intriguing topological state and transverse transport properties of ferromagnetic $FeIn_2S_4$ are revealed by first-principles calculations. The centrosymmetric $FeIn_2S_4$ shows fully spin polarized half-metal feature, mainly contributed from the Fe-$d$ orbital. The clean and linear band-crossing near Fermi level traces out nodal lines and possesses high carrier mobility with Fermi velocity of $2\times10^5$ m/s. The nodal lines protected by glide mirror plane $\widetilde{\mathcal{M}}_z$ and mirror plane $\mathcal{M}_{x-y}$ form inner nodal chains in momentum space and the glide mirror plane $\widetilde{\mathcal{M}}_{xy}$ enable the nodal lines throughout the Brillouin zone from top to bottom. The effective model and symmetry analysis further demonstrate this interesting topological state. Remarkably, the band-crossing along Γ-X high symmetry line tracing out the nodal loops are robust against SOC due to the protection of nonsymmorphic glide mirror symmetry and the SOC effect on the inner nodal chains can be negligible. Moreover, the inner nodal chain leads to the coexistence of drumhead-type surface states and closed surface Fermi arcs on the (001) surface. These nontrivial nodal loops mainly contribute to the maximum AHC of -648 (Ω·cm)$^{-1}$ and the flat energy dispersion of the 3D nodal lines are beneficial to the AHC of 192 (Ω·cm)$^{-1}$ at the energy of 15 meV below Fermi level. The maximum $\alpha_{yx}$ can reach to 2.0 (A·cm$^{-1}$K$^{-1}$) and -2.3 (A·cm$^{-1}$K$^{-1}$) at low temperature of 50 K. These findings provide a suitable magnetic semimetal candidate to investigate the transverse transport properties dominated by nodal chain.

**Acknowledgments**

This work was supported by the State Key Development Program for Basic Research of China (Nos. 2019YFA0704900, and 2022YFA1403800), the Fundamental Science Center of the National Natural Science Foundation of China (No. 52088101), the National Natural Science Foundation of China (No. 12174426), the Strategic Priority Research Program (B) of the Chinese Academy of Sciences (CAS) (No. XDB33000000), the Synergetic Extreme Condition User Facility (SECUF), and the Scientific Instrument Developing Project of CAS (No. ZDKYYQ20210003).

**Declaration of competing interest**

The authors declare that they have no known competing financial interests or personal relationships that could have appeared to influence the work reported in this paper.